\documentclass[aps,pre,superscriptaddress, floatfix,11pt]{revtex4-2}
\usepackage{graphicx}% Include figure files
\usepackage{dcolumn}% Align table columns on decimal point
\usepackage{bm}% bold math
\usepackage{amsmath}
\usepackage{amssymb,amsthm}
\usepackage{color}
\usepackage{epstopdf} 
\usepackage{subfig}
\usepackage{float,placeins}

\usepackage{epstopdf}
\usepackage{mathbbol}

\let\originalleft\left
\let\originalright\right
\renewcommand{\left}{\mathopen{}\mathclose\bgroup\originalleft}
\renewcommand{\right}{\aftergroup\egroup\originalright}

\begin{document}
\title{Improved mean-field dynamical equations are able to detect the two-steps relaxation in glassy dynamics at low temperatures}

\author{David Machado} 
\affiliation{Group of Complex Systems and Statistical Physics, Department of Theoretical Physics, University of Havana, Cuba}
\author{Roberto Mulet} 
\affiliation{Group of Complex Systems and Statistical Physics, Department of Theoretical Physics, University of Havana, Cuba}
\author{Federico Ricci-Tersenghi}
\affiliation{Dipartimento di Fisica, Sapienza Universit\`a di Roma, and CNR-Nanotec, Rome unit and INFN, Sezione di Roma1, 00185 Rome, Italy}

\date{\today}
%\pacs{05.60.Gg, 03.65.Xp, 72.10.-d, 87.15.hj}

%Quantum transport (general physics)
%Tunneling (general physics)
%Theory of electronic transport (cond mat)
%Transport dynamics (biomolecules)

\begin{abstract}
  We study the stochastic relaxation dynamics of the Ising $p$-spin model on a random graph, a well-known model with glassy dynamics at low temperatures. We introduce and discuss a new closure scheme for the master equation governing the continuous-time relaxation of the system, that translates into a set of differential equations for the evolution of local probabilities. The solution to these dynamical mean-field equations describes very well the out-of-equilibrium dynamics at high temperatures, notwithstanding the key observation that the off-equilibrium probability measure contains higher-order interaction terms, not present in the equilibrium measure. In the low-temperature regime, the solution to the dynamical mean-field equations shows the correct two-step relaxation (a typical feature of the glassy dynamics), but with a relaxation timescale too short. We propose a solution to this problem by identifying the range of energies where entropic barriers play a key role and defining a renormalized microscopic timescale for the dynamical mean-field solution. The final result perfectly matches the complex out-of-equilibrium dynamics computed through extensive Monte Carlo simulations.
\end{abstract}

\maketitle

\section{Introduction}\label{sec:Intro}

A myriad of problems from condensed matter physics \cite{Onuki}, combinatorial optimization \cite{mezard2009information}, neuroscience \cite{Amit} and machine learning \cite{hertz1991introduction} are formulated via the extremization of some function of $N$ variables. The optimization landscape is sometimes very complex, and the solution becomes difficult, or impossible, because of the presence of local attractors that slow down the exploration of the configuration space. The system is called \emph{frustrated} \cite{Binder1986, MezardParisiVirasoro}. 

The relevance of the subject is widely recognized, and some theoretical tools for the study of systems' equilibrium properties are now becoming well-established science, after years of applications in diverse contexts. The replica symmetry breaking \cite{MezardParisiVirasoro, mezard1984replica}, the cavity method \cite{MezardParisi2001, mezard2003cavity} and the Thouless-Anderson-Palmer approach \cite{thouless1977solution} allowed to tackle numerous problems. On the other hand, the progress in out-of-equilibrium situations has been considerably slower. The field lacks a general understanding of the problems, and their treatments are highly specific depending on the type of variables (continuous \cite{bouchaud1998out} or discrete \cite{derrida1987exactly}), their interactions topology (fully-connected \cite{ coolen1996DRT}, random \cite{mozeika2008dynamical} or latticed \cite{BrayDyn2D1994}), or the nature of time itself (also continuous \cite{CME-PRE} or discrete \cite{del2015dynamic}).

If we complicate the scenario by looking at problems with a known spin-glass phase in equilibrium, the available results are even more scarce. The pioneering work of Sompolinsky and Zippelius \cite{SompolinskyZippelius1981} studied the continuous-time dynamics of a soft-spin version of the Sherrington-Kirkpatrick (SK), where the variables are continuous. This case is usually modelled with a Langevin formalism as it is possible to write a differential equation directly for the soft-spin variables. Later, Sompolinsky \cite{Sompolinsky_alone_1981} attempted the construction of a consistent mean-field theory that takes account of both equilibrium and non-equilibrium spin-glass behavior. 

Motivated by the dynamical properties of spin-glasses, like the aging regime \cite{Binder1986, Alba_1986, Alba_1987}, a series of works \cite{cugliandolo1993analytical, Cugliandolo_1994, cugliandolo1995full, bouchaud1998out} pointed some inconsistencies in Sompolinsky's approach and exploited the picture of \emph{weak ergodicity breaking} \cite{Bouchaud_1992} to re-visit the problem. Besides the study of the SK model \cite{Cugliandolo_1994}, they included a simpler but illustrative model: the spherical spin-glass model with $p$-spin interactions \cite{cugliandolo1993analytical}. The theory reproduced aging and allowed for the existence of multiple time-scales. However, recently, this theory has been questioned for more complicated spherical models \cite{folena2020rethinking} and for Ising models on sparse random graphs \cite{bernaschi2020strong}.

While the spherical $p$-spin is a long-range model (fully-connected) with continuous variables, we focus our attention on the diluted ferromagnetic $p$-spin model, defined for discrete spin variables on sparse random graphs. Although in our case there is no quenched disorder in the interactions, and despite all the differences mentioned before, there is a common phenomenology. At low temperatures, both models exhibit multiple relaxation processes within the dynamic evolution, one of them with a very large characteristic time scale. This relation between traditional mean-field spin-glass models and more realistic structural glass models has been discussed in various occasions \cite{Bouchaud_1994, Franz}.

Our work exploits shared aspects between them, like the existence of multiple time scales, to give a simple mean-field theory to describe the spin glass dynamics of discrete variables on random graphs. We derive a hierarchy of approximations that allows us to consider spatial correlations with increasing accuracy. Although there is abundant literature about another important hierarchical system of equations for the glassy dynamics, known as the Generalized Mode Coupling Theory (GMCT) \cite{SzamelGMCT2003, Janssen_GMCT_2016}, it is important to emphasize here that our approaches are fundamentally different. While the GMCT does not take into account the topology of the interactions, we explicitly consider a specific random graph of interacting variables where the notions of distance and neighborhood are relevant.

This article is organized as follows. The Section \ref{sec:dyn} introduces a new closure for the master equation governing the system dynamics, written for a single instance of the interactions graph. For simplicity, we applied it to the $p$-spin ferromagnet on random regular hypergraphs, where everything is reduced to one average case equation. In Section \ref{sec:noneq_broadcasting} we generalize the work of Montanari and Semerjian \cite{Montanari2006} on dynamical phase transitions to the out-of-equilibrium scenario. The latter allows us to re-interpret the results of the dynamical theory presented in Section \ref{sec:dyn} by providing a way to compute a new time scale for our calculations. The results are compared with Monte Carlo simulations in Section \ref{sec:tscale}.

\section{Conditioned Dynamic Approximation}
\label{sec:dyn}

In its simplest form, the ferromagnetic $p$-spin model is defined by the Hamiltonian \\ $H = -\sum_{i_1, i_2, \ldots, i_p} \sigma_{i_1}\sigma_{i_2} \ldots \sigma_{i_p}$, where we have $N$ discrete-spin variables $\sigma_{i}=\pm1$. The interaction is structured in groups, called plaquettes, of exactly $p$ variables. This kind of model is usually represented using a hypergraph where the spins correspond to variable nodes (denoted by the indexes $i, j, \ldots$) and the plaquettes to factor nodes (denoted by $a, b, \ldots$). 

The continuous time dynamics of the probabilities $P^{t}(\vec{\sigma})$ of having some configuration $\vec{\sigma}=\{ \sigma_1, \sigma_2, \ldots, \sigma_N \}$ is governed by the Master Equation \cite{vanKampen92}:
\begin{equation}
\frac{d P(\vec{\sigma})}{dt} = -\sum_{k=1}^{N} r_k(\vec{\sigma}) \, P(\vec{\sigma}) + \sum_{k=1}^{N} r_k(F_k[\vec{\sigma}]) \, P(F_k[\vec{\sigma}])
 \label{eq:mas_eq_gen}
\end{equation}
where $r_k(\vec{\sigma})$ is the probability per time unit that the spin $\sigma_k$ changes to $-\sigma_k$, when the system's configuration is $\vec{\sigma}$. These are called dynamic \emph{transition rates}. The operator $F_k[\cdot]$ takes any configuration $\vec{\sigma}$ and flips the $k$-th spin to get: $\vec{\sigma}' = \{ \sigma_1, \sigma_2, \ldots, -\sigma_k, \ldots, \sigma_N \}$.

When the transition rates $r_k$ depends only on the variables that directly interact with the $k$-th node, we can choose some site $i$ and sum (\ref{eq:mas_eq_gen}) over all the configurations that keep the spin $\sigma_i$ fixed, the result is the local equation:
\begin{equation}
\frac{d P(\sigma_i)}{dt} = - \sum_{ \sigma_{\partial i}} r_i(\sigma_i, \sigma_{\partial i}) P(\sigma_i, \sigma_{\partial i}) + \sum_{ \sigma_{\partial i}} r_i(-\sigma_i, \sigma_{\partial i}) P(-\sigma_i, \sigma_{\partial i})
\label{eq:mas_eq_single}
\end{equation}
The symbol $\partial i$ represents the set of nodes that interact with $i$ according to the model's Hamiltonian. Of course, we have an equation like (\ref{eq:mas_eq_single}) for all spins in the system. But these are not the only equations we can get. We could in principle marginalize (\ref{eq:mas_eq_gen}) keeping fixed a plaquette of $p$ interacting variables and thus obtain a differential equation for the probability of a plaquette's configuration, or we could fix a variable $\sigma_i$ and all its \emph{neighborhood} $\sigma_{\partial i}$ to obtain a differential equation for $P(\sigma_i, \sigma_{\partial i})$:
\begin{eqnarray}
\frac{d}{dt} P(\sigma_a) &=& -\sum_{i \in a} \sum_{\sigma_{\partial i \setminus a}}  \Big[ r_i(\sigma_i, \sigma_{\partial i}) P(\sigma_{\partial i \setminus a}, \sigma_a) - r_i(-\sigma_i, \sigma_{\partial i}) P(\sigma_{\partial i \setminus a}, F_i[\sigma_a]) \Big] \label{eq:mas_eq_plaquette} \\
\frac{d}{dt} P(\sigma_i, \sigma_{\partial i})  &=& -r_i(\sigma_i, \sigma_{\partial i}) \, P(\sigma_i, \sigma_{\partial i}) + r_i(-\sigma_i, \sigma_{\partial i}) \, P(-\sigma_i, \sigma_{\partial i}) - \label{eq:mas_eq_exact} \\
 & &  - \sum_{b \subset \partial i}\sum_{j \in b \setminus i} \sum_{\sigma_{\partial j \setminus b}} \Big[  r_j(\sigma_j, \sigma_{\partial j}) P(\sigma_{\partial j \setminus b}, \sigma_{\partial i}, \sigma_i) - r_j(-\sigma_j, \sigma_{\partial j}) P(\sigma_{\partial j \setminus b}, F_j[\sigma_{\partial i}], \sigma_i) \Big]  \nonumber
\end{eqnarray}
Here, the indexes $i$ and $j$ represent variable nodes, while the indexes $a, b$ denote plaquettes, and $\sigma_a$ is the configuration of the variables inside the plaquette $a$. In order to lighten our notation, we preferred to think of the symbols $a$ and $b$ as sets of variables nodes. Thus, for example, $b \subset \partial i$ stands for the subset of $\partial i$ formed by the nodes \emph{inside} the plaquette $b$. 

By choosing each time a larger group of spins we can construct a hierarchical system of differential equations that we should truncate at some point. The simpler approximation that one can make is to neglect all connected correlations $C_{ij}=\langle \sigma_i \sigma_j \rangle - \langle \sigma_i  \rangle \, \langle \sigma_j \rangle$, closing the system at the level of the equation (\ref{eq:mas_eq_single}). In practice, to substitute $P(\sigma_{\partial j \setminus b}, \sigma_{\partial i}, \sigma_i)$ by $P(\sigma_i) \prod_{k \in \partial i \setminus j} P(\sigma_k)$.

In the sake of simplicity and concreteness, let us consider the ensemble of random regular hypergraphs,  where all variables $\sigma_i$ participate in the same number $c$ of plaquettes ($c$ is the node's connectivity), and all plaquettes contain exactly $p=3$ variables. These are locally tree-like structures where the typical length of the loops diverges with the system size as $\ln(N)$. 

It is important to notice that the only source of disorder in this ensemble is the presence of loops. After neglecting the correlations $C_{ij}$ and setting homogeneous initial conditions, the system can be characterized by just one differential equation.
\begin{equation}
 \frac{d \phi}{dt} = -\sum_{u=0}^{c} \binom{c}{u} r(u) \, [f_1(\phi)]^{u} \, [f_2(\phi)]^{c-u} \, \phi + \sum_{u=0}^{c} \binom{c}{u} r(u) \, [f_2(\phi)]^{u} \, [f_1(\phi)]^{c-u} \, (1 - \phi) \label{eq:me_closed_single}
\end{equation}
where $\phi(t)$ is the probability that a spin points up, $f_1(\phi) = 2 \phi (1-\phi)$ and $f_2(\phi)=\phi^{2} + (1-\phi)^{2}$. Given the structure of the Hamiltonian, we can write the transition rates in terms of a single integer: the number of unsatisfied interactions between a spin and its neighbors $u=\sum_{a \subset \partial i} \delta(\prod_{k \in a} \sigma_k, -1)$.

Within this approximation, starting the relaxation at low temperatures from the initial condition $\phi(0)=1/2$, the equation (\ref{eq:me_closed_single}) gives $\dot{\phi}(0) = 0$ and $\phi(t)=1/2$ at all times. The energy density is then easily computed as $e(t) = 0$ for all times, which obviously is very different from the real dynamics of the model.

A less trivial approximation, similar to the one employed in \cite{CME-Pspin} is the following:
\begin{eqnarray}
 P(\sigma_{a \setminus i} , \sigma_i, \sigma_{b \setminus i}) &=& P(\sigma_{b \setminus i} \mid \sigma_i, \sigma_{a \setminus i}) \, P(\sigma_a) \approx P(\sigma_{b \setminus i} \mid \sigma_i) P(\sigma_a) \nonumber \\
 P(\sigma_{a \setminus i} , \sigma_i, \sigma_{b \setminus i}) &\approx& \frac{P(\sigma_a) P(\sigma_b)}{P(\sigma_i)} \label{eq:app_CDA_1}
\end{eqnarray}
In the first line of (\ref{eq:app_CDA_1}) we approximated the conditional probability $P(\sigma_{b \setminus i} \mid \sigma_i, \sigma_{a \setminus i})$ by $P(\sigma_{b \setminus i} \mid \sigma_i)$. This means that the configuration of $\sigma_{a \setminus i}$ is irrelevant for the probability distribution of $\sigma_{b \setminus i}$ once $\sigma_i$ is given. The information of the variable at distance $d=1$ in the graph is enough.

This is reflected in the following connected correlations, computed using the conditional probability distributions:
\begin{eqnarray}
C_{\sigma_i} &\equiv& \langle  \sigma_{a \setminus i} \sigma_{b \setminus i}\rangle_{\sigma_i} - \langle   \sigma_{a \setminus i} \rangle_{\sigma_i} \langle \sigma_{b \setminus i}\rangle_{\sigma_i} \nonumber \\
C_{\sigma_i} &=& \sum_{\sigma_{a \setminus i}, \sigma_{b \setminus i}} \Big[ \prod_{k \in a \setminus i} \sigma_k \Big] \Big[ \prod_{j \in b \setminus i} \sigma_j \Big] \; P(\sigma_{a \setminus i}, \sigma_{b \setminus i} \mid \sigma_i) - \nonumber \\
& & \:\:\:\: - \Big\{ \sum_{\sigma_{a \setminus i}} P(\sigma_{a \setminus i} \mid \sigma_i) \prod_{k \in a \setminus i} \sigma_k \Big\} \Big\{ \sum_{\sigma_{b \setminus i}} P(\sigma_{b \setminus i} \mid \sigma_i) \prod_{j \in b \setminus i} \sigma_j \Big\}
\label{eq:plaq_corr}
\end{eqnarray}

In principle, $C_{\sigma}$ can be non-zero. However, (\ref{eq:app_CDA_1}) allows us to factorize the joint conditional probability distribution, giving that $P(\sigma_{a \setminus i}, \sigma_{b \setminus i} \mid \sigma_i) = P(\sigma_{a \setminus i} \mid \sigma_i) P(\sigma_{b \setminus i} \mid \sigma_i)$. This immediately leads to $C_{\sigma_i} = 0$ for both values of $\sigma_i$.

Setting the same initial condition $P(\sigma_i)=\frac{1}{2}$ for all $i$ in a random regular hypergraphs leads to a single equation for $P(\sigma_a)$, where we can drop the index $a$ and define 
\begin{equation}
\Phi\equiv \sum_{\sigma_a} P(\sigma_a) \:  \delta(\, \prod_{k \in a} \sigma_k, \: -1)
\end{equation}
The corresponding differential equation is:
\begin{equation}
 \frac{1}{p} \frac{d \Phi}{dt} = - \sum_{u=0}^{c - 1} \binom{c-1}{u} \, r(u + 1) \, \Phi^{u+1} \, (1 - \Phi)^{c - 1 - u} + \sum_{u=0}^{c - 1} \binom{c-1}{u} \, r(u) \, \Phi^{u} \, (1 - \Phi)^{c - u} 
 \label{eq:CDA_1}
\end{equation}
In what follows we call (\ref{eq:CDA_1}) the Conditional Dynamic Approximation of the first order (\emph{CDA-1}), for reasons that will be clearer latter. This equation can be numerically integrated in time to get a non-trivial relaxation of the energy density $e(t)$. 

\begin{figure}[t]
\centering
\subfloat[]{
\includegraphics[width=0.48\textwidth]{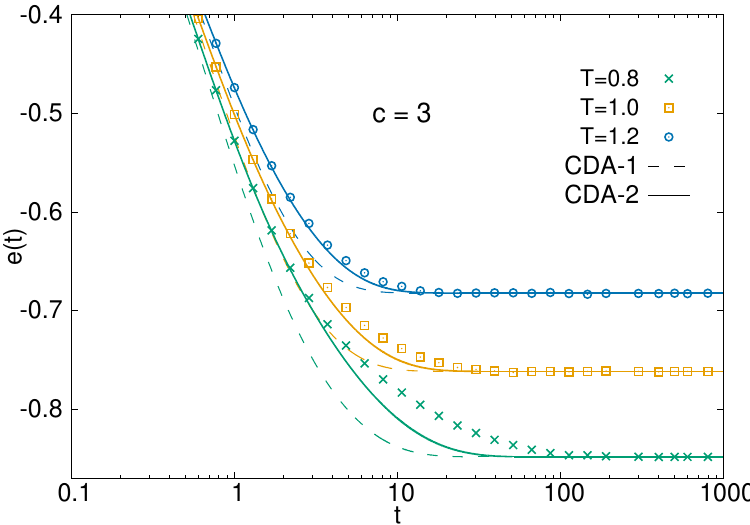} \label{fig:Pspin_KMC_e_vs_t_CDA_1_2_high_T}}
\subfloat[]{
\includegraphics[width=0.48\textwidth]{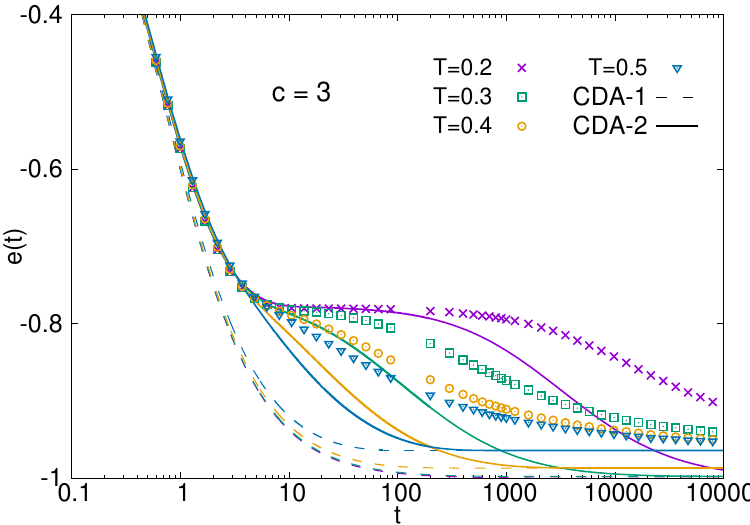} \label{fig:Pspin_KMC_e_vs_t_CDA_1_2_low_T}}
\caption{Time dependence of the energy density $e(t)$ in the $p$-spin ferromagnet with $p=3$ at several temperatures. Each point represents the average of $10^4$ Monte Carlo simulations with system size $N=10000$. Error bars are of the size of the points. The result of the numerical integration of the \emph{CDA-1} equations is represented with dashed lines, while continuous lines represent the \emph{CDA-2} approximation. Data in panel (a) and (b) are for high ($T>T_d\approx 0.51$) and low ($T<T_d$) temperatures, respectively.}
\label{fig:Pspin_KMC_e_vs_t_CDA_1_2}
\end{figure}

The results for the Glauber dynamics of this model, where we make a specific choice for the rates $r_i(\sigma_i, \sigma_{\partial_i}) = \frac{1}{2} \big( 1 - \sigma_i \tanh[\beta J \sum_{k \in \partial i} \sigma_k] \big) = \frac{1}{2} \big( 1 - \tanh[\beta J (c - 2 u)] \big)$, are compared with Monte Carlo simulations in Fig.~\ref{fig:Pspin_KMC_e_vs_t_CDA_1_2}. We represent the \emph{CDA-1} with dashed lines. For high temperatures, see Fig.~\ref{fig:Pspin_KMC_e_vs_t_CDA_1_2_high_T}, the approximation works reasonably well for the transient regime and provides the right stationary value for the energy. However, below the dynamical spin glass transition, which occurs at temperature $T_d \approx 0.51$, the \emph{CDA-1} approximation is very poor and returns a relaxation very far from the behavior measured in the simulations (see Fig.~\ref{fig:Pspin_KMC_e_vs_t_CDA_1_2_low_T}).

We are interested precisely in this two-step relaxation, which is typical of glassy dynamics at low temperatures. Given that we cannot obtain a non-zero correlation between neighboring plaquettes from the \emph{CDA-1}, we will take another step. Analogously to (\ref{eq:app_CDA_1}), to close equation (\ref{eq:mas_eq_exact}) we can write:
\begin{eqnarray}
P(\sigma_{\partial j \setminus b}, \sigma_i, \sigma_{\partial i}) &=&  P(\sigma_{\partial j \setminus b} \mid \sigma_i, \sigma_{\partial i}) \, P(\sigma_i, \sigma_{\partial i}) \approx P(\sigma_{\partial j \setminus b} \mid \sigma_j, \sigma_{b \setminus j}) P(\sigma_i, \sigma_{\partial i}) \nonumber \\
P(\sigma_{\partial j \setminus b}, \sigma_i, \sigma_{\partial i}) &\approx&  \frac{P(\sigma_{j}, \sigma_{\partial j})}{P(\sigma_i, \sigma_{b \setminus i})} \, P(\sigma_i, \sigma_{\partial i}) = \frac{P(\sigma_{j}, \sigma_{\partial j})}{\sum_{\sigma_{\partial j \setminus b}}P(\sigma_{j}, \sigma_{\partial j})} \, P(\sigma_i, \sigma_{\partial i})
\label{eq:app}
\end{eqnarray}

\begin{figure}[th]
\centering
\includegraphics[width=0.5\textwidth]{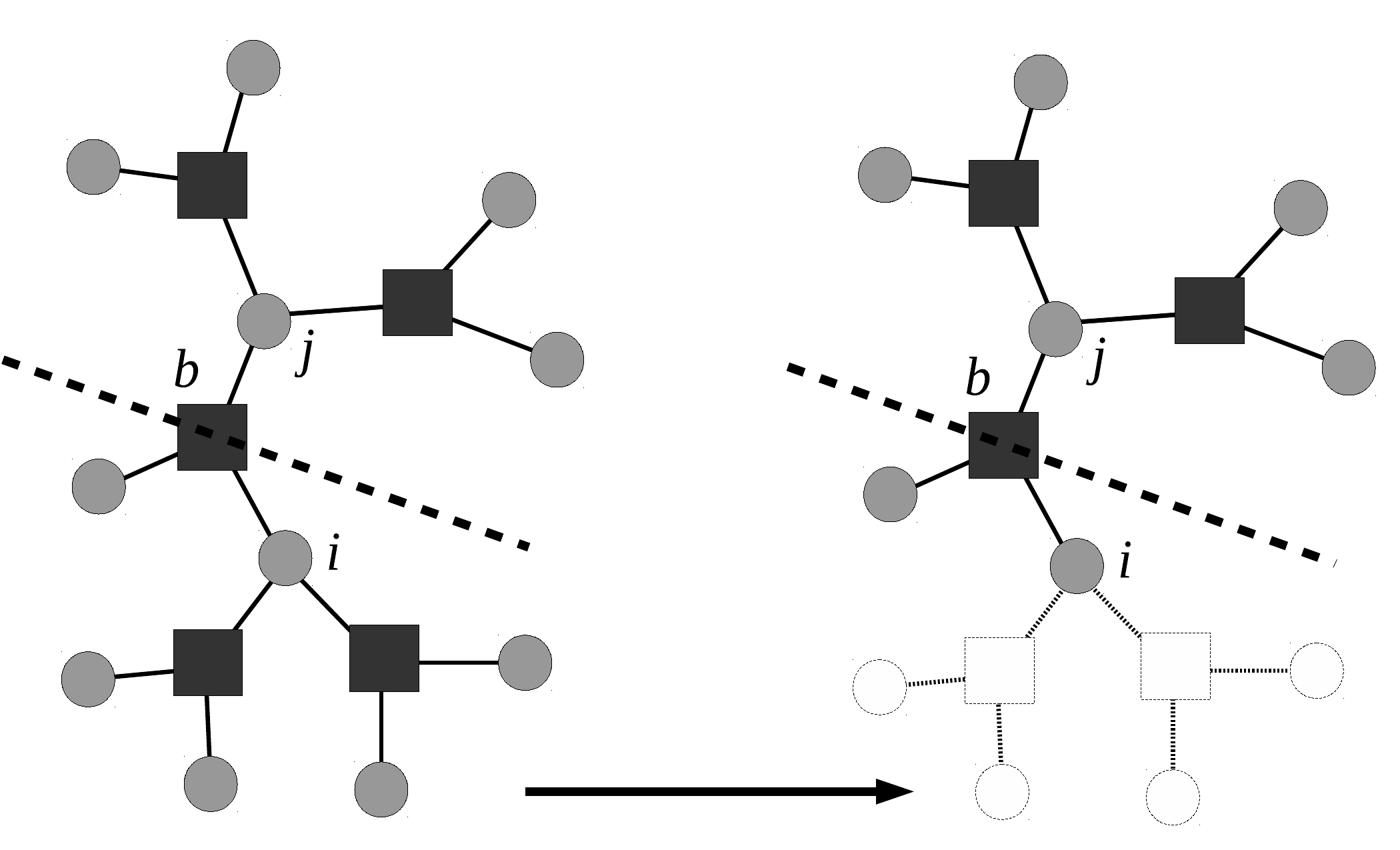}
\caption{Illustration of the Conditioned Dynamic Approximation of the second order (equation  (\ref{eq:app})). The circles represent variable nodes, and the squares represent factor nodes. The figures show the nodes involved in the conditional probability $P(\sigma_{\partial j \setminus b} \mid \sigma_i, \sigma_{\partial i})$. The variables which are above the dashed lines are inside the argument of the conditional probability ($\partial j \setminus b$), while the variable below the dashed lines are in the conditions ($\sigma_i, \sigma_{\partial i}$). The approximation consists on neglecting the effect of the nodes that are not colored in the rightmost part of the figure. As these nodes are at distance $d=2$ from the node $j$, we say that this is an approximation \textit{of the second order} in space.}
\label{fig:illustration_app}
\end{figure}

Fig.~\ref{fig:illustration_app} illustrates the meaning of equation (\ref{eq:app}), which we  call Conditioned Dynamic Approximation of the second order (\emph{CDA-2}) in what follows. The target is the conditional probability $P(\sigma_{\partial j \setminus b} \mid \sigma_i, \sigma_{\partial i})$, which involves the configuration of the nodes in the neighborhoods of $i$ and $j$.

By approximating this probability by $P(\sigma_{\partial j \setminus b} \mid \sigma_j, \sigma_{b \setminus j})$ we are neglecting the effect of the nodes which are at distance $d=3$ from $\partial j \setminus b$ (see Fig.~\ref{fig:illustration_app}), but allowing nonzero values of the connected correlations $C_{\sigma_i}$ (see Eq. \eqref{eq:plaq_corr}). Then we say that the closure is \textit{of the second order} in space. It is straightforward to generalize (\ref{eq:app}) and then write other levels of approximation: \emph{CDA-3}, \emph{CDA-4}, etcetera.

After putting Eqs.~(\ref{eq:mas_eq_exact}) and (\ref{eq:app}) together we obtain a closed system of differential equations for $P(\sigma_i, \sigma_{\partial i})$.
\begin{eqnarray}
 \frac{d}{dt} P(\sigma_i, \sigma_{\partial i})  &=& -r_i(\sigma_i, \sigma_{\partial i}) \, P(\sigma_i, \sigma_{\partial i}) + r_i(-\sigma_i, \sigma_{\partial i}) \, P(-\sigma_i, \sigma_{\partial i}) - \nonumber \\
 & &  - \sum_{b \subset \partial i}\sum_{j \in b \setminus i} \sum_{\sigma_{\partial j \setminus b}}   r_j(\sigma_j, \sigma_{\partial j}) \frac{P(\sigma_{j}, \sigma_{\partial j})}{\sum_{\sigma_{\partial j \setminus b}}P(\sigma_{j}, \sigma_{\partial j})} \, P(\sigma_i, \sigma_{\partial i}) + \nonumber \\ 
 & &  + \sum_{b \subset \partial i}\sum_{j \in b \setminus i} \sum_{\sigma_{\partial j \setminus b}} r_j(-\sigma_j, \sigma_{\partial j}) \frac{P(\sigma_{j}, \sigma_{\partial j})}{\sum_{\sigma_{\partial j \setminus b}}P(-\sigma_{j}, \sigma_{\partial j})} \, P(\sigma_i, F_{j}[\sigma_{\partial i}]) \label{eq:CDA_pspin}
\end{eqnarray}
The equations (\ref{eq:CDA_pspin}) do not contain any further assumption about the actual structure of the interactions. Actually, it is not even necessary to have exactly $p$ spins in each plaquette for these equations to be valid. The dynamical rates $r_i(\sigma_i, \sigma_{\partial i})$ can also take any the form of any function of the spin on site $i$ and its neighborhood. Thus, in the way we presented it, the \emph{CDA-2} is of general purpose.

If we consider again a random regular hypergraph with the initial condition $P(\sigma_i) = \frac{1}{2}, \forall i=1, \ldots, N$, the Eq.~\ref{eq:CDA_pspin} becomes the same for all the values of $i$. Now, all the sites are equivalent and it is possible to rewrite the probability $P(\sigma_i, \sigma_{\partial i})$ in terms of one parameter: the number of unsatisfied interactions between the central spin $\sigma_i$ and its neighbors ($u=\sum_{a \subset \partial i} \delta(\prod_{k \in a} \sigma_k, -1)$). The equation for $P(u)$ is derived in the Appendix \ref{sec:DINA} and allows a considerable simplification of the numerical computations. We concentrate here on their solution.

Fig.~\ref{fig:Pspin_KMC_e_vs_t_CDA_1_2_high_T} illustrates the behaviour of $e(t)$ for $c=3$, computing results for three specific temperatures. Together with Monte Carlo simulations (points) and the previous approximation (\emph{CDA-1}, with dashed lines), we represent the results of the \emph{CDA-2} with continuous lines. When the temperature decreases, the theoretical technique becomes slightly less accurate to describe the transient regime but it keeps predicting the steady state of the system very accurately. In all cases, the transient regime obtained from the \emph{CDA-2} is closer to the simulations than the results of the \emph{CDA-1}. 

More importantly, Fig.~\ref{fig:Pspin_KMC_e_vs_t_CDA_1_2_low_T} shows the behaviour of the model below $T_d\approx0.51$. In this case, the energy relaxation obtained from the \emph{CDA-2} approximation takes place through two different relaxation processes. This important feature of the glassy dynamics has been obtained thanks to the fact that the \emph{CDA-2} approximation takes into account the correlation between nearest neighbour energy defects. Such a correlation starts playing a key role exactly at the energy value where the first plateau develops (let us call $e_p$ this energy value). The plateau at $e_p$ becomes much longer (mind the log scale) when the temperature is decreased.

Notwithstanding the important result about the two-steps relaxation (not captured by simpler closure schemes), the \emph{CDA-2} approximation shows a too-short time scale to leave the plateau and an asymptotic energy value which depends on the temperature and is different from the energy reached in Monte Carlo numerical simulations. In fact, the final energies predicted by the \emph{CDA-2} correspond to the equilibrium paramagnetic phase, which is non-physical in the region $T<T_d$ \cite{CME-Pspin}.

We believe these two failures can be ascribed to the following approximation underlying any mean-field closure scheme: at every time during the dynamics, the average is taken over a measure assuming that all relevant configurations are easily accessible by the dynamics itself. This is in general true at high enough energies, where correlations are weak. However, for low enough energies correlations become very strong (e.g.\ the correlation between energy defects below $e_p$) and make the microscopic time scale to sample the measure larger.
When this microscopic timescale grows, the curves like $e(t)$ should be ``stretched'' and eventually, if the divergence of such a time scale takes place, the relaxation would reach a stop.

It is clear that at the plateau energy $e_p$ some relevant barrier must appear: indeed, the $T=0$ dynamics is not able to relax below $e_p$ both in Monte Carlo simulations and in the mean-field equations. These barriers are likely to have an entropic origin as discussed in detail in Ref.~\cite{bellitti2021entropic}.
Thus we can safely assume that no barrier is present for $e>e_p$ (at least along the typical trajectories followed by the relaxation dynamics) and the microscopic time scale can be set to the (conventional) unit value in such a regime.
On the contrary, barriers are present below $e_p$. But these are non-extensive barriers, that remain finite in the large $N$ limit, as witnessed by the observation that the dynamics with any $T>0$ can relax below $e_p$.

Fig.~\ref{fig:Pspin_KMC_e_vs_t_CDA_1_2_low_T} also shows that Monte Carlo simulations relax to an asymptotic energy $e_d\approx -0.96$ strictly larger than $-1$. This is the so-called dynamical threshold energy. At this energy value, barriers become extensive, i.e.\ diverge in the large $N$ limit, and the relaxation gets stuck. In the next Section, we are going to compute the dynamical threshold energy, extending the classical computation of the point-to-set correlation function to the out-of-equilibrium regime.

\section{Non-equilibrium dynamical transition}\label{sec:noneq_broadcasting}

Models on random graphs that manifest frustration have been shown to exhibit purely dynamical phase transitions \cite{Bouchaud_1994,Franz}, where the ergodicity is broken but the free energy remains analytic. In such a complex situation, the work of Montanari and Semerjian \cite{Montanari2006} gives a simple method to detect the occurrence of the transition.

In short, the method reduces to look at the relation of a variable $\sigma_0$ in the system, with the variables $\lbrace \sigma_l \rbrace$ at a distance $l$ of $\sigma_0$. For a given configuration of $\lbrace \sigma_l \rbrace$, it is possible to use a message passing technique \cite{pearl2014probabilistic} to compute the expected value $m_0^{l}$ of the spin $\sigma_0$. As we explain in the Appendix \ref{sec:PtSCorr}, a set of self-consistent equations allows to obtain the probability distribution of expected values, $Q^{l}(m_0^{l})$, as a function of the distance $l$. Then the magnitudes to measure are the point-to-set correlation $\mathcal{C}^{l} = \int dm \, Q^{l} (m) \, m$ and the corresponding length $l(\epsilon) = \text{min} \big\lbrace l: \mathcal{C}^{l} < \epsilon \big\rbrace$.

The point-to-set correlation length $l(\epsilon)$ diverges exactly at $T_d$, regardless of the value chosen for $\epsilon$. Thus, the problem of detecting the presence of a dynamic phase transition is simplified to the determination of the divergence of a single magnitude \cite{Montanari2006}.

Here, we extend the calculation to consider local measures that are completely out of equilibrium. Let us assume that from a dynamical computation we are able to obtain the local probabilities:
\begin{eqnarray}
 P(S_{a \setminus i}, S_{b \setminus i} \mid \sigma_i) &=& \sum_{\sigma_{a \setminus i}} \sum_{\sigma_{a \setminus i}} P(\sigma_{a \setminus i}, \sigma_{b \setminus i} \mid \sigma_i) \, \delta(S_{a \setminus i} , \prod_{j \in \partial a \setminus i} \sigma_j) \, \delta(S_{b \setminus i}, \prod_{k \in \partial b \setminus i} \sigma_k) \label{eq:local_prob}
\end{eqnarray}
where $\delta(x, y)$ stands for the Kronecker delta on $x$ and $y$.

With the information contained in (\ref{eq:local_prob}) we can write the set of self-consistent equations for the cavity probability distributions $Q^{l}(\mu)$ and $Q^{l}(\hat{\mu})$ (see Appendix \ref{sec:PtSCorr}): 
\begin{eqnarray}
 Q^{l}(\mu) &=& \sum_{S_{1}, \ldots, S_{c-1}} \hat{\pi}(S_{1}, \ldots, S_{c-1}) \int \Big[\prod_{i=1}^{c-1} d\hat{\mu}_i \; \hat{Q}^{l, S_i}(\hat{\mu}_i) \Big] \; \delta\Big( \mathcal{F}_1\big[\mu,  \lbrace \hat{\mu}_i \rbrace_{i=1}^{c-1} \big] \Big) \label{eq:up_m_mod} \\
 \hat{Q}^{l, S}(\hat{\mu}) &=& 2^{-p+2} \!\!\!\!\!\! \sum_{\sigma_1, \ldots, \sigma_{p-1}} \delta(S, \, \prod_{j=1}^{p-1} \sigma_j) \int \Big[\prod_{j=1}^{p-1} d\mu_j \; Q^{l-1}(\sigma_j \, \mu_j) \Big] \; \delta\Big(\mathcal{F}_2^{S}\big[ \hat{\mu},  \lbrace \mu_j \rbrace_{j=1}^{p-1} \big] \Big) \label{eq:up_hat_m_2}
\end{eqnarray}
where we considered again the case where $P(\sigma_i) = \frac{1}{2}$ for all $i=1, \ldots, N$ and have defined $\hat{\pi}(S_{1}, \ldots, S_{c-1}) \equiv P(S_1, \ldots, S_{c-1} \mid \sigma = 1)$. The Dirac delta functions in these equations enforce the relations:
\begin{eqnarray}
 \mathcal{F}_1\big[\mu,  \lbrace \hat{\mu}_i \rbrace_{i=1}^{c-1} \big] &=& \frac{1 + \mu}{2} - \frac{\prod_{i=1}^{c-1} \big(1 + \hat{\mu}_i \big)}{\sum_{\sigma} \Big[\prod_{i=1}^{c-1} \big(1 + \sigma \hat{\mu}_i \big) \Big]}  \label{eq:up_m_f1} \\
 \mathcal{F}_2^{S}\big[ \hat{\mu},  \lbrace \mu_j \rbrace_{j=1}^{p-1} \big] &=& \hat{\mu} - S \, \langle S \rangle \prod_{j=1}^{p-1} \mu_{j}  \label{eq:up_m_f2}
\end{eqnarray}
where $\langle S \rangle = \sum_{S} \, S \, \hat{\pi}(S)$

\begin{figure}[t]
\centering
\subfloat[]{
\includegraphics[width=0.48\textwidth]{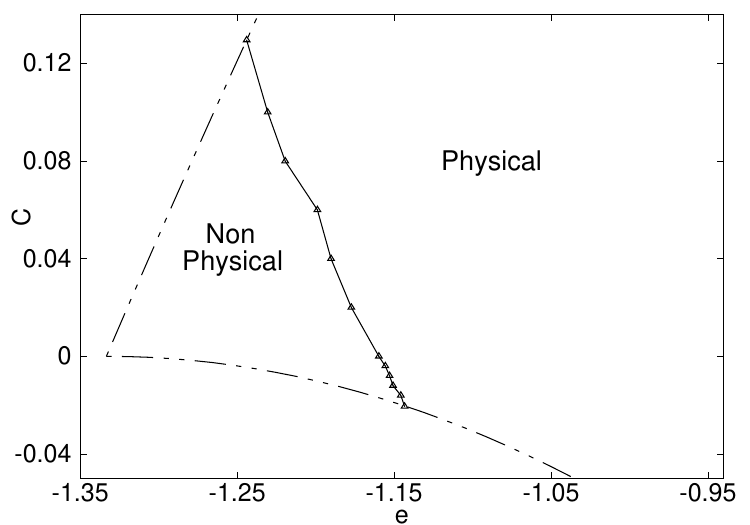}\label{fig:Phase_diagram_C_vs_e}}
\subfloat[]{
\includegraphics[width=0.48\textwidth]{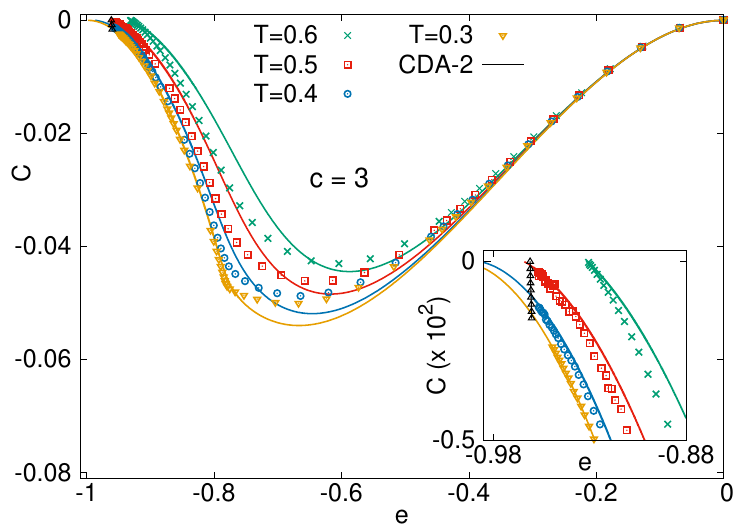} \label{fig:Pspin_KMC_C_vs_e_CDA_MC}}
\caption{(a) Phase diagram in the $(e,C)$ plane for $p=3$ and $c=4$. The continuous line with points marks the dynamical critical line where the relaxation time scale diverges. The dashed lines are physical bounds on $e$ and $C$, derived using conditions $0 \leq \hat{\pi}(S_1, S_2) \leq 1$. (b) Parametric plots $(e(t),C(t))$ for the $p$-spin ferromagnet with $p=3$ and $c=3$ at low temperatures. Each point represents the average over $10^4$ Monte Carlo simulations with system size $N=10000$. Error bars are of the size of the points. The result of the numeric integration of the \emph{CDA-2} equations is shown in continuous lines.}
\label{fig:Phase_diagram_C_vs_e_with_CDA}
\end{figure}

Now we can explore the solutions to Eqs.~(\ref{eq:up_m_mod}) and (\ref{eq:up_hat_m_2}) to locate the divergence of the corresponding dynamic point-to-set correlation. For the sake of simplicity, we write the approximation $\hat{\pi}(S_{1}, \ldots, S_{c-1}) \approx \displaystyle \prod_{i=1}^{c-2}  \hat{\pi}(S_i, S_{i+1}) \:/ \: \displaystyle \prod_{i=2}^{c-2} \hat{\pi}(S_i)$ for any local probability $\hat{\pi}$. The ``pair'' probabilities $\hat{\pi}(S_i, S_{i+1})$ can be exactly written in terms of two parameters: the energy density $e$ and the conditional correlation between neighboring plaquettes $C_{\sigma_i = 1} \equiv C$ (see Eq. \eqref{eq:plaq_corr}). Fig.~\ref{fig:Phase_diagram_C_vs_e} shows the position in the plane $(e,C)$ of the divergence of the point-to-set correlation, i.e.\ the physical time scale, obtained from Eqs.~(\ref{eq:up_m_mod}) and (\ref{eq:up_hat_m_2}).

In Fig.~\ref{fig:Pspin_KMC_C_vs_e_CDA_MC} we report the same data shown in Section \ref{sec:dyn}, but as a  parametric plot of $(e(t),C(t))$. In this representation the absolute time becomes irrelevant and we observe a good similarity between the Monte Carlo data (points) and the analytical results based on the \emph{CDA-2} approximation (lines). This is a strong indication that a proper redefinition of the microscopic time scale in the dynamical mean-field equations could eventually provide a solution very close to the physical one.

The inset in Fig.~\ref{fig:Pspin_KMC_C_vs_e_CDA_MC} is a zoom on the region reached at very long times. The black line with points marks the place where the point-to-set correlation length diverges. As Ref.~\cite{Montanari2006} points out, this is to be associated with the divergence of the physical time scale. Indeed the Monte Carlo simulation is not able to go to the left of this line, into the non-ergodic region (see Fig.~\ref{fig:Phase_diagram_C_vs_e}), and it seems to converge very closely to the dynamical critical line at large times. On the contrary, the \emph{CDA-2} approximation returns trajectories that enter into the non-ergodic region, evidently ignoring the dynamical phase transition.

\section{Setting the proper time scale in the dynamical mean-field equations}{\label{sec:tscale}}

As shown in Fig.~\ref{fig:Pspin_KMC_e_vs_t_CDA_1_2_low_T}, at low temperatures, the \emph{CDA-2} approximation shows up two failures compared to the physical dynamics obtained through Monte Carlo simulations: (i) a too-fast relaxation below the plateau energy $e_p$ and (ii) a convergence to energy values below the dynamical threshold energy $e_d$ (see the inset of Fig.~\ref{fig:Pspin_KMC_C_vs_e_CDA_MC}), in the non-ergodic region, which should not be accessible on finite time scales. We cannot correct (ii) by going to higher order closures of the CDA hierarchy because they will not capture the divergence of the point-to-set correlation, a key ingredient for the existence of $e_d$. On the other hand, we do not expect the \emph{CDA-3} and higher levels to significantly improve (i) as to justify the necessary efforts.

Nonetheless, in the $(e,C)$ plane, the \emph{CDA-2} approximated trajectories follow very closely the true dynamics, showing non-equilibrium correlations (i.e. $C\neq 0$).
So, we foresee the possibility of achieving a very good matching between the true and the \emph{CDA-2} approximated dynamics by just redefining the microscopic timescale in the dynamical mean-field equations.

The reason beyond this sort of time-reparametrization comes from the following argument: within the mean-field approximation, at every time, one takes the average over a particular probability distribution that should correspond to the configurations that the system is likely to visit at that time. As long as this probability distribution is well ergodic, the hypothesis that the required average can be taken in a short time is reasonable. However, when barriers come into play and the probability distribution to sample is no longer well ergodic, this could, in turn, correspond to a longer effective time scale. Only if this longer time scale is used in the solution to the mean-field dynamical equations, then there is a chance of describing the actual dynamics. The same argument can be brought to the extreme consequences when the \emph{CDA-2} approximated trajectory approaches the dynamical critical line, where the relaxation time scale diverges because in the probability measure the ergodicity breaks down.

As discussed above, the microscopic time scale does not need to be renormalized for energies large enough ($e>e_p$) because in this region there are no barriers. Below $e_p$ barriers arise and we assume that the exploration of the accessible configurations is slowed down by a factor $\exp(\Delta_S)$. The entropic barrier $\Delta_S$ is constant with respect to time and does not depend on the temperature.
Finally, approaching the dynamical phase transition, the microscopic time scale must diverge and we assume a simple power law divergence $(e-e_d)^{-\gamma}$.

So, the simpler ansatz for the effective time scale is the following
\begin{equation}
\hat{\tau}(e) =
\left\{
\begin{array}{ll}
1 & e_p < e \\
e^{\Delta_S} \big((e_p-e_d) \, / \, (e - e_d)\big)^{\gamma} \quad\ & e_d < e < e_p \\
\infty & e < e_d
\end{array}
\right.
\label{eq:newt}
\end{equation}
This ansatz depends on two important energy values, $e_p$ and $e_d$, that we now discuss in detail, and two parameters, $\Delta_S$ and $\gamma$, that will be adjusted to match the actual physical dynamics.

On the one hand, $e_p$ is the energy density of the intermediate plateau, which does not depend on the temperature, and thus can be computed by solving the \textit{CDA-2} equations at $T=0$.
On the other hand, $e_d$ marks the frontier between ergodic and non-ergodic configurations as computed in the Section \ref{sec:noneq_broadcasting} (see Fig.~\ref{fig:Phase_diagram_C_vs_e}). As already discussed, $e_d$ marks the place where the physical time scale diverges. Below $e_d$ the system relaxation would proceed by activation processes, which are ignored in the present approach. 

The expression in Eq.~\eqref{eq:newt} for the effective time scale depends on four parameters: $e_p$, $e_d$, $\Delta_S$ and $\gamma$. Two of them, $e_p$ and $e_d$, can be computed analytically as explained above. The remaining two, $\Delta_S$ and $\gamma$, will be fixed by best fitting the data from Monte Carlo simulations. We stress, however, that their values do not depend on the temperature and so we have to fix just two parameters to describe the relaxation in the entire low temperature phase.

\begin{figure}[t]
\subfloat[]{
\includegraphics[width=0.48\textwidth]{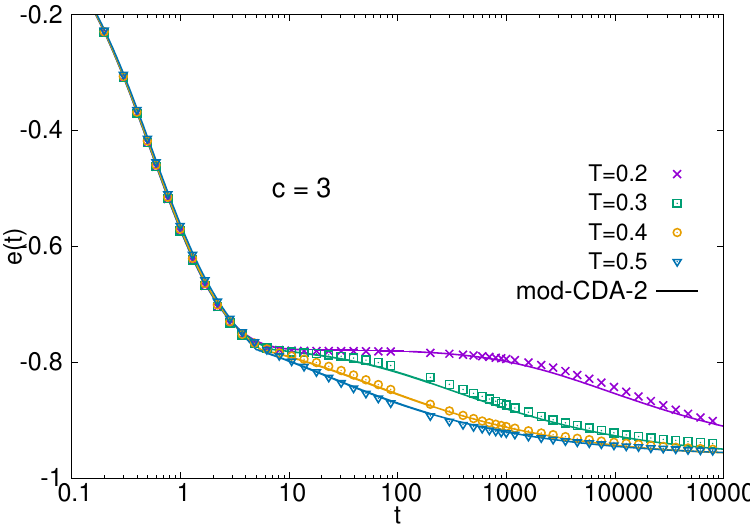}\label{fig:Pspin_KMC_e_vs_newt_c_3}}
\subfloat[]{
\includegraphics[width=0.48\textwidth]{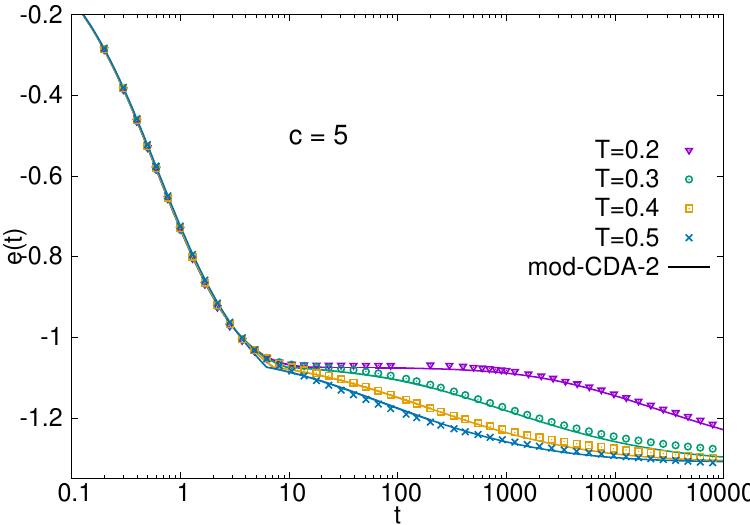} \label{fig:Pspin_KMC_e_vs_newt_c_5}}
\caption{Time dependence of the energy density $e(t)$ in the $p$-spin ferromagnet with $p=3$ at low temperatures. Each point represents the average over $10^4$ Monte Carlo simulations with system size $N=10000$. Error bars are of the size of the points. The integration of the \emph{CDA-2} equations is performed using the effective time scale defined in Eq.~(\ref{eq:newt}) and reported with he continuous lines. In (a) the graph mean degree is $c=3$ and best fitting parameters are $\Delta_S = 0.15$ and $\gamma=2$, while in (b) we have $c=5$, $\Delta_S=0.17$ and $\gamma=2$.}
\label{fig:Pspin_KMC_e_vs_newt}
\end{figure}

Fig.~\ref{fig:Pspin_KMC_e_vs_newt} shows a comparison between Monte Carlo simulations and the dynamics derived from the \emph{CDA-2} equations when the proper time scales $\hat\tau(e)$ is taken into account.
We observe a very good agreement between the actual physical dynamics and the predictions from the \emph{CDA-2} approximation. The results are presented for two different connectivities ($c=3$ and $c=5$) in order to show their robustness. The best-fitting parameters are reported in the caption. While $\Delta_S$ seems to change a little bit with $c$, the $\gamma$ parameter is very stable ($\gamma=2$ fits perfectly the data).

This $\gamma$ value lies within the bounds derived in Ref.~\cite{Montanari2006, Montanari2006ineqs}. In particular, one could be tempted to compare it to the results of the Monte Carlo simulations reported in Ref.~\cite{Montanari2006ineqs} ($\gamma_{\text{\tiny MC}}\approx 3.2$). However, the key observation is that the computation of Ref.~\cite{Montanari2006ineqs} is made on the equilibrium measure, while here we are studying an out-of-equilibrium  process. The latter is likely to make smart choices to relax towards the lowest energy $e_d$ and it is not surprising that along these smart relaxation paths, the divergence of the time scale approaching the ergodicity-breaking transition is less severe.

\section{Conclusions} \label{sec:concl}

While the glassy dynamics of continuous and unbounded variables interacting through a fully connected topology has been solved a long time ago \cite{cugliandolo1993analytical, Cugliandolo_1994, cugliandolo1995full} (although maybe not in all the aspects \cite{folena2020rethinking,folena2023weak}) the glassy dynamics of other types of models is very hard to approximate at the analytical level. In this work, we have made an important step forward in the study of Ising models defined on sparse random graphs. By using a continuous time description, we have been able to reproduce faithfully the relaxation of the energy in the low-temperature regime of the diluted Ising $p$-spin model, where such a relaxation takes place following a two steps process (see Fig.~\ref{fig:Pspin_KMC_e_vs_newt}).

Our solution takes into account two crucial aspects of the dynamics, which were not fully considered in previous works. On the one hand, we derive the Conditioned Dynamical Approximation, which is a closure scheme for the master equation that keeps track of the correlation between nearest neighbour energy defects.
This correlation is related to the presence of local barriers that need to be crossed in order to proceed with the relaxation to lower energies.
On the other hand, we realized that the mean-field equations are valid only under the assumption that all configurations with given mean-field parameters are well sampled: this can happen only on a time scale that grows approaching the ergodicity breaking transition. Following this idea, we have computed the point-to-set correlation \cite{Montanari2006} on the out-of-equilibrium measure predicted by the mean-field approximation, and defined a proper effective time scale $\hat\tau(e)$ for the mean-field evolution, see Eq.\eqref{eq:newt}.

The ideas above provide a very satisfying result, with the solution to the \textit{CDA-2} dynamical mean-field equations matching perfectly the complex energy relaxation measured in Monte Carlo simulations.
Nonetheless, there are many possible extensions of our approach which is worth pursuing in the near future.
For example, observables depending on two times may show aging, even when one-time observables (like the energy) have reached their stationary value \cite{bouchaud1998out}. Checking the quality of the mean-field approximation presented in this work for those observables is certainly very useful.
Moreover, we foresee the possibility to extend the present approach to other interaction topologies, like non-regular random graphs, or even graphs with short loops. The extension to other models (e.g. vector spin models) is very open and requires some analytical work in addition.
A quite different, but equally interesting, developing direction is to consider dynamics different from the Glauber one, eventually dynamics not satisfying detailed balance. This very broad class of processes are fundamental, for example, in the description of smart search algorithms to solve combinatorial optimization problems. Our methodology could be used to predict possible regions of divergence of the dynamical time scale and to estimate the relaxation near the algorithmic thresholds.

\begin{acknowledgments}
This research has been supported by ICSC--Centro Nazionale di Ricerca in High Performance Computing, Big Data, and Quantum Computing funded by European Union--NextGenerationEU.
\end{acknowledgments}

\bibliographystyle{unsrt}
\bibliography{ref_cvme}
\pagebreak

\appendix

\section{Average case version of the CDA-2}\label{sec:DINA}

In this section we show how to write an average case version of the \emph{CDA-2} for the dynamics on a random regular hypergraph with homogeneous initial conditions: $P(\sigma_i) = \frac{1}{2}$ for all $i=1, \ldots, N$. Then, we indicate how this particular version of the single-instance \emph{CDA-2} is related with the average case equations presented in \cite{Semerjian_dyn_2004}.

Let us remember the \emph{CDA-2} as written in the main text:

\begin{eqnarray}
 \frac{d}{dt} P(\sigma_i, \sigma_{\partial i})  &=& -r_i(\sigma_i, \sigma_{\partial i}) \, P(\sigma_i, \sigma_{\partial i}) + r_i(-\sigma_i, \sigma_{\partial i}) \, P(-\sigma_i, \sigma_{\partial i}) - \nonumber \\
 & &  - \sum_{b \subset \partial i}\sum_{j \in b \setminus i} \sum_{\sigma_{\partial j \setminus b}}   r_j(\sigma_j, \sigma_{\partial j}) \frac{P(\sigma_{j}, \sigma_{\partial j})}{\sum_{\sigma_{\partial j \setminus b}}P(\sigma_{j}, \sigma_{\partial j})} \, P(\sigma_i, \sigma_{\partial i}) + \nonumber \\ 
 & &  + \sum_{b \subset \partial i}\sum_{j \in b \setminus i} \sum_{\sigma_{\partial j \setminus b}} r_j(-\sigma_j, \sigma_{\partial j}) \frac{P(\sigma_{j}, \sigma_{\partial j})}{\sum_{\sigma_{\partial j \setminus b}}P(-\sigma_{j}, \sigma_{\partial j})} \, P(\sigma_i, F_{j}[\sigma_{\partial i}]) \label{eq:CDA_pspin_appendix}
\end{eqnarray}

We will assume that our system of spins is defined over a regular hypergraph, where every node has the same number $c$ of neighbors. In that case, if the initial conditions for the probabilities are independent of the site, all probabilities will be governed by identical equations.

Now, it is possible to write everythin in terms of one important parameter: the number $u=\sum_{a \subset \partial i} \delta(\prod_{k \in a} \sigma_k, -1)$ of unsatisfied interactions between $\sigma_i$ and its neighbors. The equations can be re-casted as:

\begin{eqnarray}
 \frac{d}{dt} P(u) &=& -r(u) \, P(u) + r(c - u) \, P(c - u) - \nonumber \\
 & & \!\!\!\!\!\!\!\!\!\!\!\!\!\!\!\! - (p-1) u \Big[\sum_{\hat{u}=0}^{c - 1} \binom{c - 1}{\hat{u}} P(\hat{u} + 1) \Big]^{-1} \sum_{\hat{u}=0}^{c - 1}   \binom{c - 1}{\hat{u}} r(\hat{u} + 1) P(\hat{u} + 1) P(u) \nonumber \\
 & & \!\!\!\!\!\!\!\!\!\!\!\!\!\!\!\! + (p-1) u \Big[\sum_{\hat{u}=0}^{c - 1} \binom{c - 1}{\hat{u}} P(\hat{u}) \Big]^{-1} \sum_{\hat{u}=0}^{c - 1}   \binom{c - 1}{\hat{u}} r(\hat{u}) P(\hat{u}) P(u - 1) \nonumber \\
 & & \!\!\!\!\!\!\!\!\!\!\!\!\!\!\!\! - (p-1) (c - u) \Big[\sum_{\hat{u}=0}^{c - 1} \binom{c - 1}{\hat{u}} P(\hat{u}) \Big]^{-1} \sum_{\hat{u}=0}^{c - 1}   \binom{c - 1}{\hat{u}} r(\hat{u}) P(\hat{u}) P(u) \label{eq:mas_eq_app} \\
 & & \!\!\!\!\!\!\!\!\!\!\!\!\!\!\!\! + (p-1) (c - u) \Big[\sum_{\hat{u}=0}^{c - 1} \binom{c - 1}{\hat{u}} P(\hat{u} + 1) \Big]^{-1} \sum_{\hat{u}=0}^{c - 1}   \binom{c - 1}{\hat{u}} r(\hat{u} + 1) P(\hat{u} + 1) P(u + 1) \nonumber
\end{eqnarray}

In the sake of simplicity, we will introduce the probabilities:
\begin{equation}
\hat{P}(u) = \binom{c}{u} P(u)
 \label{eq:def_P_hat}
\end{equation}

The Eq.~(\ref{eq:mas_eq_app}) can be expressed in terms of this $\hat{P}$ probabilities making some small re-arrangements. The sums over the variable $\hat{u}$, that go from $\hat{u}=0$ to $\hat{u} = c - 1$, should be modified like in the following example:

\begin{eqnarray}
 \sum_{\hat{u}=0}^{c - 1} \binom{c - 1}{\hat{u}} P(\hat{u} + 1) &=& \sum_{\hat{u}=0}^{c - 1} \frac{(c - 1)!}{\hat{u}! \, (c - 1 - \hat{u})!} P(\hat{u} + 1) \nonumber \\
  \sum_{\hat{u}=0}^{c - 1} \binom{c - 1}{\hat{u}} P(\hat{u} + 1) &=&  \sum_{\hat{u}=0}^{c - 1} \frac{(c - 1)!}{\hat{u}! \, (c - 1 - \hat{u})!} \, \frac{(\hat{u} + 1)! \, (c - \hat{u} - 1)!}{c!} \hat{P}(\hat{u} + 1) \nonumber \\
  \sum_{\hat{u}=0}^{c - 1} \binom{c - 1}{\hat{u}} P(\hat{u} + 1) &=& \frac{1}{c} \sum_{\hat{u}=0}^{c - 1} (\hat{u} + 1) \hat{P}(\hat{u} + 1) = \frac{1}{c} \sum_{u=0}^{c} u \, \hat{P}(u) \nonumber\\
  \sum_{\hat{u}=0}^{c - 1} \binom{c - 1}{\hat{u}} P(\hat{u} + 1) &=& \frac{1}{c} \, \langle u \rangle \label{eq:sum_unsat}
\end{eqnarray}

The Eq.~(\ref{eq:def_P_hat}) was used in the second line of Eq.~(\ref{eq:sum_unsat}) to write $P(\hat{u} + 1)$ in terms of $\hat{P}(\hat{u} + 1)$. In the last line, the notation $\langle \cdot \rangle \equiv \sum_{u=0}^{c} [\, \cdot \, ] P(u)$ was introduced.

The following identities can be derived similarly as in Eq.~(\ref{eq:sum_unsat}):

\begin{eqnarray}
 \sum_{\hat{u}=0}^{c - 1} \binom{c - 1}{\hat{u}} P(\hat{u}) &=& \frac{1}{c} \, \langle c - u \rangle \label{eq:sum_sat} \\
 \sum_{\hat{u}=0}^{c - 1} \binom{c - 1}{\hat{u}} r(\hat{u} + 1) P(\hat{u} + 1) &=&  \frac{1}{c} \, \langle \, u \, r(u) \, \rangle \label{eq:sum_r_unsat} \\
 \sum_{\hat{u}=0}^{c - 1} \binom{c - 1}{\hat{u}} r(\hat{u}) P(\hat{u})  &=& \frac{1}{c} \, \langle \, (c - u) \, r(u) \, \rangle \label{eq:sum_r_sat}
\end{eqnarray}

Putting (\ref{eq:sum_unsat}), (\ref{eq:sum_sat}), (\ref{eq:sum_r_unsat}) and (\ref{eq:sum_r_sat}) into (\ref{eq:mas_eq_app}), the differential equation becomes:

\begin{eqnarray}
 \frac{d}{dt} P(u) &=& -r(u) \, P(u) + r(c - u) \, P(c - u) - \nonumber \\
 & & - (p-1) \frac{\langle \, u \, r(u) \, \rangle}{\langle u \rangle} \, \big[ u \, P(u) - (c - u) P(u + 1) \big]  \nonumber \\
 & & - (p-1) \frac{\langle \, (c - u) \, r(u) \, \rangle}{\langle c - u \rangle} \, \big[ (c - u) \, P(u) - u P(u - 1) \big] \label{eq:Semerjian_prev}
\end{eqnarray}

Our final average case equation is obtained multiplying (\ref{eq:Semerjian_prev}) by $\binom{c}{u}$ and making use of (\ref{eq:def_P_hat}) and the relations:

\begin{eqnarray}
 \binom{c}{u} \, (c - u) \, P(u + 1) = (u + 1) \hat{P}(u + 1) \\
 \binom{c}{u} \, u \, P(u - 1) = (c - u + 1) \hat{P}(u - 1)
\end{eqnarray}

The result is:

\begin{eqnarray}
 \frac{d}{dt} \hat{P}(u) &=& -r(u) \, \hat{P}( u) + r(c - u) \, \hat{P}(c - u) - \nonumber \\
 & & - (p-1) \frac{\langle \, u \, r(u) \, \rangle}{\langle u \rangle} \, \big[ u \, \hat{P}(u) - (u+1) \hat{P}(u + 1) \big]  \nonumber \\
 & & - (p-1) \frac{\langle \, (c - u) \, r(u) \, \rangle}{\langle c - u \rangle} \, \big[ (c - u) \, \hat{P}(u) - (c - u + 1) \hat{P}(u - 1) \big] \label{eq:Semerjian_pspin}
\end{eqnarray}

Although these equations are a simplified version of the single-instance \emph{CDA-2} for the p-spin model, and the equations presented in \cite{Semerjian_dyn_2004} are written for the pairwise interactions of Ising spins in a Bethe lattice, it is possible to recognize some similarities. 

In fact, we could depart from the pairwise version of the \emph{CDA-2}:

\begin{eqnarray}
 \frac{d}{dt} P(\sigma_i, \sigma_{\partial i}) &=& -r_i(\sigma_i, \sigma_{\partial i}) \, P(\sigma_i, \sigma_{\partial i}) + r_i(-\sigma_i, \sigma_{\partial i}) \, P(-\sigma_i, \sigma_{\partial i}) - \nonumber \\
 & & - \sum_{j \in \partial i} \sum_{\sigma_{\partial j \setminus i}}   r_j(\sigma_j, \sigma_{\partial j}) \frac{P(\sigma_{j}, \sigma_{\partial j})}{\sum_{\sigma_{\partial j \setminus i}}P(\sigma_{j}, \sigma_{\partial j})} \, P(\sigma_i, \sigma_{\partial i}) + \nonumber \\ 
 & & + \sum_{j \in \partial i} \sum_{\sigma_{\partial j \setminus i}} r_j(-\sigma_j, \sigma_{\partial j}) \frac{P(\sigma_{j}, \sigma_{\partial j})}{\sum_{\sigma_{\partial j \setminus i}}P(-\sigma_{j}, \sigma_{\partial j})} \, P(\sigma_i, F_{j}[\sigma_{\partial i}]) \big\} \label{eq:mas_eq_with_app}
\end{eqnarray}

and then follow a very similar path. The only difference is that now we set the initial conditions $P(\sigma_i) = p$, allowing $p \neq \frac{1}{2}$. This introduces another relevant parameter: the value of the central spin $\sigma_i\equiv \sigma$. Apart from that, the procedure is analogous and we can find:

\begin{eqnarray}
 \frac{d}{dt} \hat{P}(\sigma, u) &=& -r(u) \, \hat{P}(\sigma, u) + r(c - u) \, \hat{P}(-\sigma, c - u) - \nonumber \\
 & & - \frac{\langle \, u \, r(u) \, \rangle_{-\sigma}}{\langle u \rangle_{-\sigma}} \, \big[ u \, \hat{P}(\sigma, u) - (u+1) \hat{P}(\sigma, u + 1) \big]  \nonumber \\
 & & - \frac{\langle \, (c - u) \, r(u) \, \rangle_{\sigma}}{\langle c - u \rangle_{\sigma}} \, \big[ (c - u) \, \hat{P}(\sigma, u) - (c - u + 1) \hat{P}(\sigma, u - 1) \big] \label{eq:Semerjian_pairwise}
\end{eqnarray}
 where $\langle \cdot \rangle_{\sigma} \equiv \sum_{u=0}^{c} [\, \cdot \, ] P(\sigma, u)$.
 
 With this, we have re-obtained the Dynamic Independent-Neighbor Approximation (DINA), as presented in \cite{Semerjian_dyn_2004}. We could then think of the \emph{CDA-2} as a more general and single-instance version of the DINA. A version that can be easily adapted to other models and graphs' architectures.
 
\section{Broadcasting process}\label{sec:PtSCorr}

Let us start in a system with a single spin variable $\sigma_0$, called root, and perform the following \emph{broadcasting} process:

\begin{enumerate}
 \item Set some positive integer $l$
 \item Create $c$ groups of $p - 1$ variable nodes.
 \item Select the values of the $p-1$ variables in each group according certain conditional probability \\ $\pi(\sigma_1, \ldots, \sigma_{p-1} \mid \sigma_0)$.
 \item Connect each group to $\sigma_0$, thus forming $c$ factor nodes, each one of order $p$.
 \item For the newly created variable nodes, repeat the steps $2$, $3$ and $4$, but instead of creating $c$ factor nodes, create $c-1$.
 \item Repeat $5$ until the last created nodes are at distance $d=l$ of the central node with value $\sigma_0$
\end{enumerate}

At the end, we obtain a graph of connected nodes: the root $\sigma_0$, and $l$ generations of nodes at distances $d=1, 2, 3, \ldots, l$ from the root. The last generation is known as the \textit{border} of the graph. Now we can ask ourselves: \textit{If we repeat this process many times for a given $l$ and keep track of the configurations of the border, can we recover the value of $\sigma_0$?}

The quantity we would like to study is the expected value $m_0^{l}$ of the central variable for a given configuration of the border at distance $l$. Then, one possible way of answering the previous question is to define some self-consistent equations for the probability distribution $Q_{\sigma_0, \pi}^{l}(m_0^{l})$ of having the expected value $m_0^{l}$ according to the border, if the border itself was generated following a broadcasting process done with root value $\sigma_0$ and using some measure $\pi(\sigma_1, \ldots, \sigma_{p-1} \mid \sigma_0)$. This is actually a somewhat involved object, and we can understand it by imagining the following stochastic process: from a given $\sigma_0$ and using some $\pi$, generate a border at distance $l$ from $\sigma_0$, and using only the information from that border, compute somehow the expected value $m_0^{l}$ of the central variable. Repeat this to get a collection of values of $m_0^{l}$ and then define $Q_{\sigma_0, \pi}^{l}(m_0^{l})$ as the probability of obtaining a given $m_0^{l}$ when doing many realizations of the stochastic process.

In order to be able of recovering the original value of the central variable, $Q_{\sigma_0, \pi}^{l}(m_0^{l})$ must have some non-trivial shape. Now, what is a non-trivial shape? One has to define some quantity to measure, and that is the expected point-to-set correlation (following the definition in \cite{Montanari2006}):

\begin{eqnarray}
C_{PS}^{l, \pi} &=& \big\langle \, \sigma_0 m_{0}^{l} \, \big\rangle_{l, \pi} - \big\langle \, \sigma_0 \, \big\rangle_{l, \pi} \big\langle \,  m_{0}^{l} \, \big\rangle_{l, \pi} \nonumber \\
C_{PS}^{l, \pi} &=&  \int dm_0^{l} \sum_{\sigma_0} P(\sigma_0) Q_{\sigma_0, \pi}^{l}(m_0^{l}) \, \sigma_0 \,  m_0^{l} - \Big( \sum_{\sigma_0} P(\sigma_0) \sigma_0 \Big) \Big( \int dm_0^{l} \sum_{\sigma_0} Q_{\sigma_0, \pi}^{l}(m_0^{l}) \,  m_0^{l} \Big)
 \label{eq:PtS_corr}
\end{eqnarray}

If we draw $\sigma_0=\pm 1$ from the uniform distribution $P(\sigma_0) = 1 / 2$, the second term in (\ref{eq:PtS_corr}) vanishes and it is not difficult to see that:

\begin{equation}
C_{PS}^{l, \pi} = \int dm_0^{l} \, Q_{+, \pi}^{l} (m_0^{l}) \, m_0^{l}
 \label{eq:PtS_corr_uniform}
\end{equation}
where $Q_{+, \pi}^{l}(m_0^{l})\equiv Q_{\sigma_0=1, \pi}^{l}(m_0^{l})$, $Q_{-, \pi}^{l}(m_0^{l})\equiv Q_{\sigma_0=-1, \pi}^{l}(m_0^{l})$ and it is necessary to use the fact that when $P(\sigma_0)$ is uniform the symmetry of the problem leads to the relation $Q_{+, \pi}^{l}(m_0^{l})\equiv Q_{-, \pi}^{l}(-m_0^{l})$.

An important point to solve the problem is to introduce two new parameters. They will be defined in broadcasting processes that are slightly different from the original one. The first parameter will be $\mu_0^{l}$, which is the expected value of the root defined in a broadcasting process where also at the first step one generates $c-1$ new factor nodes, instead of $c$. We will denote by $Q_{\sigma_0, \pi}^{l}(\mu_0^{l})$ the probability density of having some value of $\mu_0^{l}$. The second parameter is the expected value $\hat{\mu}_0^{l}$ taken in a broadcasting process where at the first step one generates only one factor node, and the corresponding probability density will be denoted as $\hat{Q}_{\sigma_0, \pi}^{l}(\hat{\mu}_0^{l})$. 

For those distribution the relations $Q_{+, \pi}^{l}(\mu_0^{l}) \equiv Q_{-, \pi}^{l}(-\mu_0^{l})$ and $\hat{Q}_{+, \pi}^{l}(\hat{\mu}_0^{l}) \equiv \hat{Q}_{-, \pi}^{l}(-\hat{\mu}_0^{l})$ must also hold. One has then two different distributions $Q_{+, \pi}^{l}(\mu_0^{l})$ and $\hat{Q}_{+, \pi}^{l}(\hat{\mu}_0^{l})$, that in what follows will be denoted simply as $Q_{\pi}^{l}(\mu)$ and $\hat{Q}_{\pi}^{l}(\hat{\mu})$. We will only need then the probability $\pi(\sigma_1, \ldots, \sigma_{p-1} \mid \sigma_0=1)$, which will also be re-denoted as $\pi(\sigma_1, \ldots, \sigma_{p-1})$.

The advantage of introducing $\mu$ and $\hat{\mu}$ is that one can compute the corresponding distributions using the following iterative equations:

\begin{eqnarray}
 Q_{\pi}^{l}(\mu) &=& \int \Big[\prod_{i=1}^{c-1} d\hat{\mu}_i \; \hat{Q}_{\pi}^{l}(\hat{\mu}_i) \Big] \; \delta\Big( \mathcal{F}_1\big[\mu,  \lbrace \hat{\mu}_i \rbrace_{i=1}^{c-1} \big] \Big) \label{eq:up_m} \\
 \hat{Q}_{\pi}^{l}(\hat{\mu}) &=& \sum_{\sigma_1, \ldots, \sigma_{p-1}} \pi(\sigma_1, \ldots, \sigma_{p-1}) \int \Big[\prod_{j=1}^{p-1} dm_j \; Q_{\pi}^{l-1}(\sigma_j \, \mu_j) \Big] \; \delta\Big(\mathcal{F}_2\big[ \hat{\mu},  \lbrace \mu_j \rbrace_{j=1}^{p-1} \big] \Big) \label{eq:up_hat_m}
\end{eqnarray}
with the relations
\begin{eqnarray}
 \mathcal{F}_1\big[\mu,  \lbrace \hat{\mu}_i \rbrace_{i=1}^{c-1} \big] &=& \frac{1 + \mu}{2} - \frac{\prod_{i=1}^{c-1} \big(1 + \hat{\mu}_i \big)}{\sum_{\sigma} \Big[\prod_{i=1}^{c-1} \big(1 + \sigma \hat{\mu}_i \big) \Big]}  \label{eq:up_m_f1_app} \\
 \mathcal{F}_2\big[ \hat{\mu},  \lbrace \mu_j \rbrace_{j=1}^{p-1} \big] &=& \hat{\mu} - S \, \langle S \rangle \prod_{j=1}^{p-1} \mu_{j}  \label{eq:up_m_f2_app}
\end{eqnarray}
where

\begin{equation}
S=\prod_{i=1}^{p-1} \sigma_i \:\:\:\:\:\:\:\: \text{and} \:\:\:\:\:\:\:\: \langle S\rangle = \sum_{\sigma_i, \ldots, \sigma_{p-1}} \Big( \prod_{i=1}^{p-1} \sigma_i \Big) \, \pi(\sigma_i, \ldots, \sigma_{p-1}) 
\end{equation}

When $\pi(\sigma_1, \ldots, \sigma_{p-1})$ is taken according the Boltzmann distribution, if a non-trivial fixed point of (\ref{eq:up_m}) and (\ref{eq:up_hat_m}) exists, it corresponds to a non-trivial fixed point of the cavity method at the 1-RSB level with parameter $x=1$ \cite{Montanari2006}. In such non-trivial fixed points, $C_{PS}^{l, \pi} \not\to 0$ when $l \to \infty$. We can define a correlation length as the minimum distance $l$ at which the value of $C_{PS}^{l, \pi}$ is smaller that certain parameter:

\begin{equation}
 l^{\ast}(\epsilon) = \text{min} \big\lbrace l: C_{PS}^{l, \pi} < \epsilon \big\rbrace \label{eq:PtSCorr_length}  
\end{equation}

The point-to-set correlation length $l^{\ast}(\epsilon)$ must then diverge at $T_d$ the temperature of the dynamic spin-glass transition.

To compute $C_{PS}^{l, \pi}$ we could use first a populations dynamics algorithm to obtain the distributions $Q_{\pi}^{l}(\mu)$ and $\hat{Q}_{\pi}^{l}(\hat{\mu})$. With the latter it is possible to find the probability density $Q_{\pi}^{l}(m)$ of the original broadcasting problem in a random regular hypergraph with:

\begin{eqnarray}
 Q_{\pi}^{l}(m) &=& \int \Big[\prod_{i=1}^{c} d\hat{\mu}_i \; \hat{Q}_{\pi}^{l}(\hat{\mu}_i) \Big] \; \delta\Big( \mathcal{F}_3\big[m,  \lbrace \hat{\mu}_i \rbrace_{i=1}^{c} \big] \Big) \label{eq:up_m_normal} \\
 \mathcal{F}_3\big[m,  \lbrace \hat{\mu}_i \rbrace_{i=1}^{c} \big] &=& \frac{1 + m}{2} - \frac{\prod_{i=1}^{c} \big(1 + \hat{\mu}_i \big)}{\sum_{\sigma} \Big[\prod_{i=1}^{c} \big(1 + \sigma \hat{\mu}_i \big) \Big]}  \label{eq:up_m_f3}
\end{eqnarray}

And then, with $Q_{\pi}^{l}(m)$ one can directly compute (\ref{eq:PtS_corr_uniform}). However, if we just want to know where the point-to-set correlation length diverges, it is enough to define the cavity point-to-set correlation $\tilde{C}_{PS}^{l, \pi} = \int d\mu \, Q_{\pi}^{l} (\mu) \, \mu$ and the corresponding length $
 \tilde{l}^{\ast}(\epsilon) = \text{min} \big\lbrace l: \tilde{C}_{PS}^{l, \pi} < \epsilon \big\rbrace$.

\begin{figure}[H]
\centering
  \subfloat[]{
	   \centering
	   \includegraphics[width=0.45\textwidth]{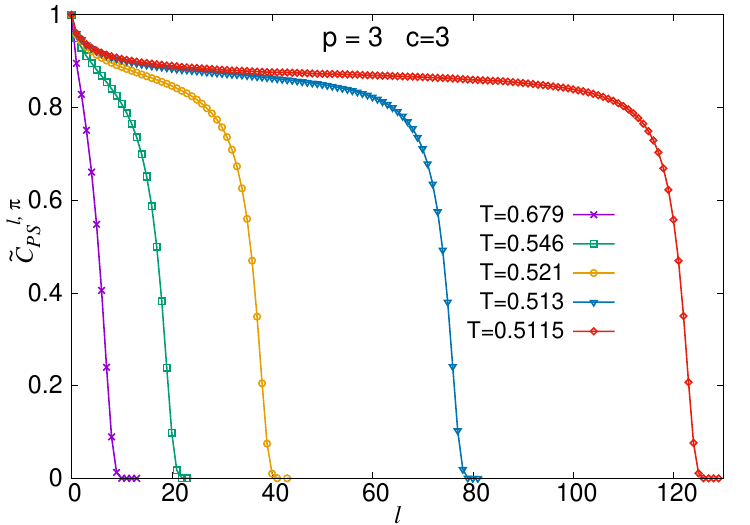}}
  \subfloat[]{
	   \centering
	   \includegraphics[width=0.45\textwidth]{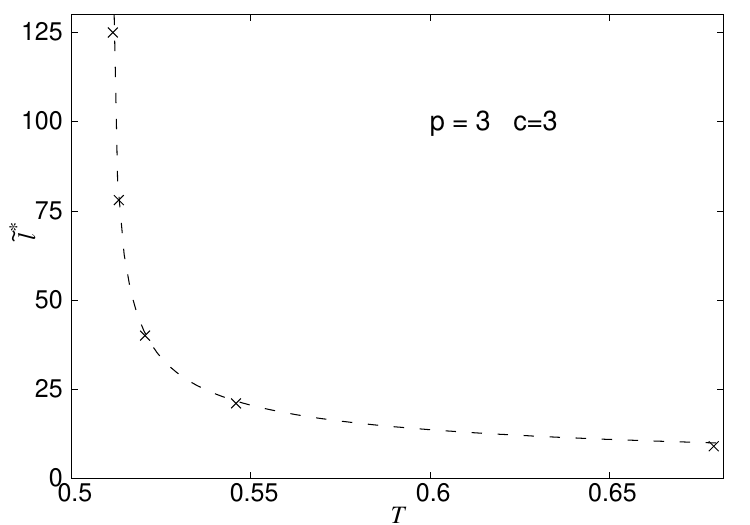}}
	   
\caption{Cavity point-to-set correlations in the $p$-spin ferromagnet defined over random regular hypergraphs with $c=3$ and $p=3$. \textbf{a)} Dependence of the cavity point-to-set correlation on the distance  for several temperatures. \textbf{b)} Dependence on the temperature of the cavity point-to-set correlation length computed with small parameter $\epsilon=0.05$. A fit to the law $\tilde{l}^{\ast}=A / (T - Td)^{1/2}$ (shown in dashed lines) gives $T_d \approx 0.51$}
\label{fig:PtSCorr_RRG_cr1_0}
\end{figure}

Fig.~\ref{fig:PtSCorr_RRG_cr1_0}a shows that near $T_d$ the value of $\tilde{l}^{\ast}(\epsilon)$ becomes larger, and Fig.~\ref{fig:PtSCorr_RRG_cr1_0}b shows a non-linear fit of these values according to the law $A / (T - T_d)^{1/2}$, which gives $T_d \approx 0.51$. This is the value that has been reported in the literature for the transition temperature of the $p$-spin ferromagnet defined over random regular hypergraphs with $c=3$ and $p=3$

\end{document}